\documentclass[prl,showpacs,showkeys,preprintnumbers,email]{revtex4}
\usepackage{epsfig}
\usepackage{color}

\begin{document}

\title{Digital Processing in Tunneling Spectroscopy}

\author{E.\,M. Dizhur\/,
        A.\,V.\,Fedorov}
\address{Institute for High Pressure Physics of the RAS, 142190, 
 Troitsk, Moscow Reg., Russia}

\date{\today}

\begin{abstract}

An alternative approach to detect very weak singularities on the
characteristics of a tunnel diode is proposed in which the numerical
differential filtering is applied directly to measured current versus
voltage dependence instead of the modulation technique commonly used
with this purpose. The gains and looses of the both approaches in the
particular case of tunneling investigations of semiconductors under
pressure are discussed. The corresponding circuitry and mathematical
routines are presented.

\end{abstract}


\email{dizhur@ns.hppi.troitsk.ru}
\maketitle

\section{Introduction} 
One of the modern topics of condensed matter physics is concerned
with the studies of the effects of many-particle interactions on
electronic transport phenomena.

Many-particle interactions (change-correlation and/or electron-phonon)
may manifest themselves in the tunneling characteristics either due to
the variation of the barrier shape or due to the variations of the tunneling
probability induced by including additional tunneling channels~\cite{bib:Theory}. E.g., the
interaction of the electrons with LO-phonons produces specific singularities
in the tunneling spectra of the GaAs-based junctions at biases $\pm 36.5 meV$,
corresponding to the energy of LO-phonon in GaAs. Inter-electron interaction
in the plasma of a semiconductor is responsible for the appearance of so called
Zero Bias Anomaly (ZBA) that looks like a peak of the tunneling resitance at
small bias voltages. While magnetoresistive measurements supply the
experimental data only for filled subbands of dimensional quantization,
the tunneling may provide information for initially empty subbands also.
It was shown recently ~\cite{bib:HPSP10}, that the tunnel junction Al/Delta(Si)-GaAs is a
convenient object for investigations of the pressure influence on many-particle
effects in the 2D electron system, and the possible metal-insulator
type transition was predicted in near-surface $\delta$-doped layer in
Al/GaAs($\delta$-Si) structure under pressure of about 2 GPa.

The observation and the experimental study of this phenomenon implied the
necessity of performing tunnel measurements at very large resistivities of
the samples aiming to detect very weak singularities on the current-voltage
characteristics of a tunnel diode.

In the tunneling spectroscopy a current through the tunnel junction is measured
versus the applied bias voltage. The commonly used way of retrieving tunnel
spectra is based on performing measurements using a modulation technique. In
such a method the sample is feeded with a DC bias with an addition of small AC
component. The resulting current may be expressed as~\cite{bib:modul}
\begin{equation}
I(V + a \sin(\omega t))\approx I(V)+ a \frac{dI}{dV}\Big\vert_V \sin(\omega t)
- \frac{1}{2} a^2 \frac{dI^2}{dV^2} \Big\vert_V \sin^2(\omega t) + \cdots 
\end{equation}
and corresponding derivatives may be approximated as
\begin{equation}
\frac{dI}{dV} \Big\vert_V \sim \sin(\omega t + \phi_1);
\frac{dI^2}{dV^2}\Big\vert_V \sim \sin(2 \omega t + \phi_2).
\end{equation}

This gives a possibility to reveal very slight singularities, containing a
valuable physical information. The straightforward application of a computer
to such a measuring procedure is to use the usual modulation technique for
analog processing of the signal and then digitize and store the obtained data
for future numerical treatment.

This way is the best since it is the simplest one, but this is not the case
when dealing with tunnel junctions of very high resistivity. From the
experimental point of view the reason for this is that the analog data
averaging is performed as a rule with a lock-in amplifier that should be tuned
precisely enough to the phase of the signal and should have rather large time
constant for better resolution.

The first obstacle is that $ I(V)$ characteristic of a semiconductor tunnel
junction is highly non-linear so that the phase shift varies during the bias
sweep. Next, the active resistance of the bias grows rapidly (by several orders
of magnitude) with pressure so that the capacitive constituent of the impedance
becomes significant enough to make the phase shift even more important and the
phase tuning more difficult.  And the last but not least is the interference
with industrial pulse noises which also increases as the resistance of the
junction increases. The two former problems could be, in principle, partially
solved by lowering the frequency of the modulation but this way would result
in the increase of the averaging time of the lock-in amplifier and would also
provide the additional problems with the pulse noises as the probability of
their appearance during the measurements increases with the duration of the
accumulation time, thus making good measurements practically impossible.

That is why we used instead fully DC measurements of the tunnel current and
tried to compensate the dynamical range deterioration using proper mathematical
treatment of the digital readings. 

\section{Circuitry details}
Fig.1 shows a schematic view of the sample under investigations and its
equivalent circuit. The latter includes the resistances of the metal leads
to the gate and the resitances of the two halves of the $\delta$-layer. The
typical values are 
$C \sim 1$\, nF, $R_{Me} \sim 10$\, Ohm, $R_{\delta} \sim 5$\, kOhm $\div 10$\, MOhm
and $R_{tun} \sim 0.2 \div 300$\, MOhm at helium temperatures in the pressure
range from zero up to $2$\, GPa respectively.

Fig.2 shows schematically the circuitry used both to feed the sample with given bias and to
obtain the measered current. Its main feature is a usage of Analog Device
Ultralow  Input Current Operational Amplifier  AD549KH with the input bias current
lower than $60$ fA. The first stage provides the maintenance of the bias
voltage applied to the tunnel junction equal to the given output voltage from
the DAC, since the voltage drop along the metal leads and along the $\delta$-doped
layer is negligible even at $\delta$-layer resistances up to 1 GOhm. The second
stage is an ordinary current-to-voltage transformer.

A measurement cycle consisted of the following steps:

1. Performing multiple readings at each given bias voltage.

2. Elimination of the obvious under- and overshots in the array of these readings
with a median filtering and evaluating the dispersion of the rest data set.

3. Using a non-equidistant bias sweep, providing finer steps
at biases where {\it a priori} known features of interest should exist - near
zero (for ZBA) and near $\pm 36.5$\,mV (for LO-phonon energies).

The measured $I(V)$ curves were subjected to the mathematical treatment to
obtain finer features of the tunneling current.

\section{Treatment} 
In a common modulation technique, the compromise between resolution and
signal-to-noise ratio is achieved by tuning the amplitude of the modulation
voltage.

Our approach includes two-stage smoothing $I(V)$ presented in the tabular form
using a smoothing cubic spline.
At the first stage the smoothing parameter was chosen so that to obtain the
compromise between the noise and the clearness of the many-particle
singularities at the second derivative of the spline using weight function
inversly proportional to the previousely obtained dispersions and "an eye
judgement". This procedure is illustrated by Fig.3.
At the second stage further smoothing was performed to smooth out the fine
features, thus obtaining a background.
The latter reveals the energy position of the subbands of the dimensional
quantization, and the difference between the first-stage smoothing and the
background being the contribution of the many-particle interaction
({\it cf.} Fig.4). 
The corresponding fragment of the mathematical routines written in
MATLAB are presented in the Appendix~A.

\section{Conclusion} 
The comparison of the approaches based on a common modulation
technique and the present one shows that the results obtained for the sample
eligible for the both of them are consistent with each other. The traditional
modulation technique gives about twice better resolution at relatively low
impedances, nevertheless fails at highly resistive samples, while the present
technique is applicable even for the samples with the resistance in GOhm range.

The present approach has enhanced stability with respect to the industrial
noises due to the possibility to eliminate overshots at the accumulation stage
while the traditional one is very vulnarable especially at long accumulation times.

\begin{acknowledgments} 
The authors acknowledge financial support by Russian Foundation for Basic
Researches and by the grant of Physical Sciences Department of RAS. 
\end{acknowledgments}


\section{APPENDIX A} 
This is a fragment of the simplified MATLAB procedure used to process
experimental data.

\begin{verbatim}
...
%	s123 is a three-column array containing bias voltage, current and the dispersion
%	Sorting by bias in ascending order
[xx,ii]=sort(s123(:,1));yy=s123(ii,2);ss=s123(ii,3);
%	Over- and undershots elimination
mss=mean(ss);ii=find(ss>4*mss);xx(ii)=[];yy(ii)=[];ss(ii)=[];
X=xx/max(abs(xx));Y=yy/max(abs(yy)); 	%Normalization 
%	Weights 
W0=min(ss+1e-6)./(ss+1e-6)+1;
%	The smoothing is less important at biases of interest
%       and more important when obtaining the background.
%       This is accounted for by weights W1 and W2 respectively.
IW=find(xx>-0.045&xx<-0.025|abs(xx)<0.005|xx<0.045&xx>0.025);
W1=W0;W2=W0;W1(IW)=max(W0);W2(IW)= min(W0);
...	% Graphic output for the visual audit is skipped
%	Build the smoothing cubic spline for I(V)
sp=csaps(X,Y,1-2^(-6),[],W1);
%	and calculate the logarithmic derivative of the conductivity
D1=fnval(fnder(sp,1),X); D2=fnval(fnder(sp,2),X); Y1=D2./D1;
%	Tabulate on the equidistant grid for convenience
xx2=linspace(xx(1),xx(end),length(xx));
X2=xx2/max(abs(xx2));
Y2=interp1(X,Y1,X2);
%	Build the "strongly" smoothing cubic spline for the background
sp2=csaps(X2,Y2,1-2^(-6)/3,[],W2);
%	Calculate the background and the many-particle singularities
BG=fnval(sp2,X2); Sings= Y2-BG ;
%	The background minima correspond to the subbands position
SubBands=X2(find(diff(sign(diff(BG)))==2)+1);
\end{verbatim}

\newpage

\begin{figure}[h]
\includegraphics[height=8cm,keepaspectratio]{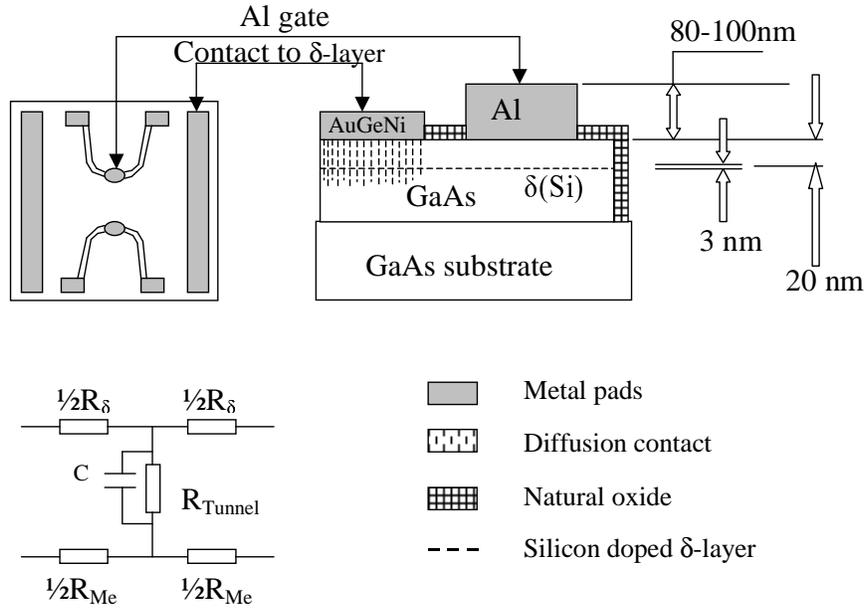}
\caption{Plain view, schematic structure and equivalent circuit
of the sample Z1B7 ~\cite{bib:sample}.} 
\label{1}
\end{figure}

\begin{figure}[h]
\includegraphics[height=8cm,keepaspectratio]{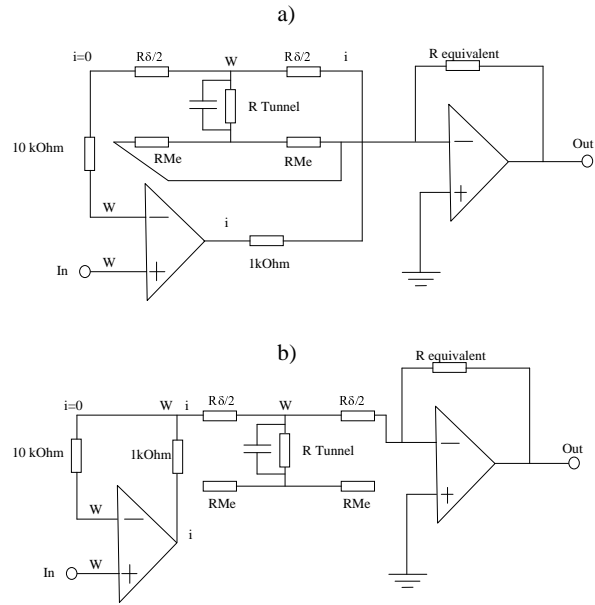}
\caption{The principal schematics of the measurement circuit.
The controls for initial balancing and commutation are omitted for clarity.
a) and b) are the variants for $I(V)$ tunneling and lateral measurements
respectively.} 
\label{2}
\end{figure}

\newpage

\begin{figure}[th]
\includegraphics[height=8cm,keepaspectratio]{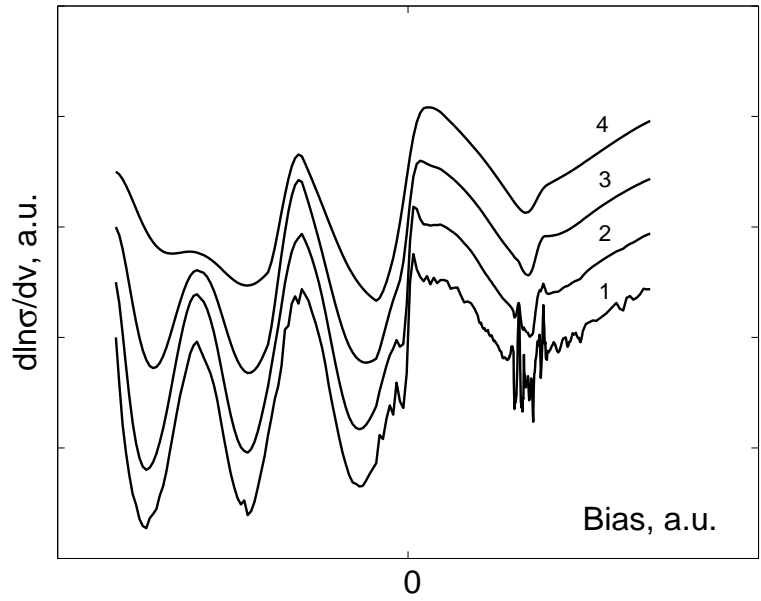}
\caption{Shape of the $\frac{dI^2}{dV^2}=\frac{d\ln\sigma}{dV}$
for Z1B7 at zero pressure and at $T=4.2$\,K with different smoothing
rates. The curves are enumerated in ascending smoothing rate. The
curve 2 is considered to be the best.} 
\label{3}
\end{figure}

\begin{figure}[bh]
\includegraphics[height=9cm,keepaspectratio]{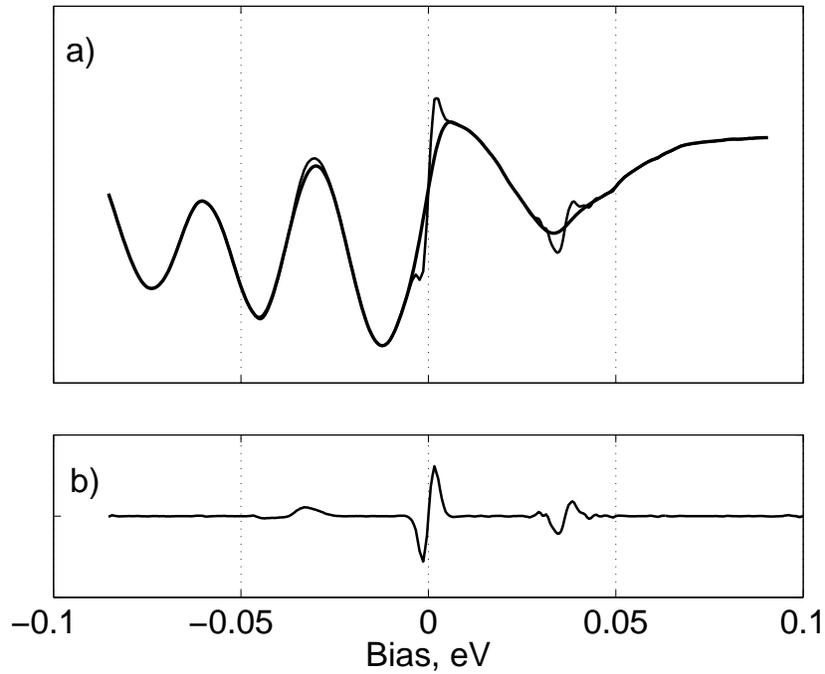}
\caption{a) Further smoothing of the curve 2 from Fig.3 results in
a background. b) The difference of the above curves gives the
contribution due to the many-particle interactions - ZBA and
the two LO-phonon structures whose different shapes are
physically meaningful.} 
\label{4}
\end{figure}

\end{document}